\documentclass[amssymb,pra,twocolumn,notitlepage,superscriptaddress]{revtex4-2}
\pdfoutput=1
\usepackage{times}
\usepackage{amsmath}
\usepackage{ulem}
\usepackage{graphicx}
\usepackage{color}

\usepackage[colorlinks=true, letterpaper=ture, pdfstartview=FitV, linkcolor=blue, citecolor=blue, urlcolor=blue]{hyperref}

\begin{document}
\title{Full-counting statistics of particle distribution on a digital quantum computer}
\author{Yun-Zhuo Fan}
\affiliation{Key Laboratory of Atomic and Subatomic Structure and Quantum Control (Ministry of Education), and  School of Physics, South China Normal University, Guangzhou 510006, China}
\author{Dan-Bo Zhang}
\email{dbzhang@m.scnu.edu.cn}
\affiliation{Key Laboratory of Atomic and Subatomic Structure and Quantum Control (Ministry of Education), and  School of Physics, South China Normal University, Guangzhou 510006, China}
\affiliation{Guangdong Provincial Key Laboratory of Quantum Engineering and Quantum Materials,  Guangdong-Hong Kong Joint Laboratory of Quantum Matter, and Frontier Research Institute for Physics,\\  South China Normal University, Guangzhou 510006, China}
\date{\today}
	
\begin{abstract}
Full-counting statistics (FCS) provides a powerful framework to access the statistical information of a system from the characteristic function. However, applications of FCS for generic interacting quantum systems often be hindered by the intrinsic difficulty of classical simulation of quantum many-body problems. Here, we propose a quantum algorithm for FCS that can obtain both the particle distribution and cumulants of interacting systems. The algorithm evaluates the characteristic functions by quantum computing and then extracts the distribution and cumulants with classical post-processing. With digital signal processing  theory, we analyze the dependency of accuracy with the number of sampling points for the characteristic functions. We show that the desired number of sampling points for accurate FCS can be reduced by filtering some components of the quantum state that are not of interest. By numeral simulation, we demonstrate FCS of domain walls for the mixed Ising model. The algorithm suggests an avenue for studying full-counting statistics on quantum computers.
\end{abstract}

\maketitle
	
\section{introduction}
Statistical information including both distribution and the cumulants is undoubtedly necessary to characterize and understand broad physical systems of interest. The full-counting statistics, initiated by Levitov and Lesovik~\cite{levitov1992charge}, provides a general framework to access all statistical information of a system by deriving the distribution and cumulants from the characteristic function. As a powerful method, the FCS has been widely applied in mesoscopic physics~\cite{beenakker2001counting,bagrets2003full,belzig2005full,gustavsson2006counting,matthiesen2014full,yu2016full}, ultracold atomic systems~\cite{nunnenkamp2006full,rath2010full,malossi2014full,lovas2017full,devillard2020full} and critical dynamics~\cite{del2018universal,cui2020experimentally}. Besides, the FCS can be used for detecting the entanglement entropy for many-body quantum systems~\cite{klich2009quantum,song2011entanglement,song2012bipartite,calabrese2012exact,jiang2022fermion}.
    
The framework of FCS, while generally applicable, relies heavily on the efficient calculation of the characteristic function. This turns to be feasible for non-interacting quantum systems, for which the characteristic function can often be analytically obtained and sequentially the distribution can be derived by Fourier transformation. However, for generic quantum many-body systems with interaction, evaluation of the characteristic function meets a challenge due to the exponential growth of Hilbert space. The difficulty is intrinsic, although under some specific circumstances perturbation methods~\cite{vzonda2016perturbation,sulejmanpasic2018aspects,madsen2021quantum,acevedo2022quantum,mitarai2023perturbation} or classical simulation by quantum Monte Carlo may work~\cite{foulkes2001quantum,pieper2001quantum,troyer2005computational,rubtsov2005continuous,carlson2015quantum}.
    
The intrinsic difficulty of solving quantum systems classically can be overcome if one can refer to a well-controlled quantum system to simulate it, as proposed by Feynman~\cite{feynman2018simulating}. For a target quantum system, the quantum computer directly simulates it by preparing a quantum state describing the system and then extracting observables by performing measurements on the quantum state. The static properties can be simulated by preparing the system's eigenstates~\cite{babbush2014adiabatic,kandala2017hardware,wang2019accelerated,nakanishi2019subspace,motta2023quantum} or thermal states at finite temperature~\cite{rousochatzakis2019quantum,sen2020nuclear,bertini2021finite,sun2021quantum,nishino2021finite}, while the dynamical properties rely on performing Hamiltonian evolution from an initial state~\cite{trotter1959product,suzuki1976generalized,lloyd1996universal,campbell2019random,chen2021concentration,yang2022improved}. Quantum computing provides a new path for us to study the FCS for generic interacting quantum systems by directly measuring the characteristic function for a prepared state on the quantum processor. Such a concept of studying FCS by quantum computing is very natural. While there are some works that have proposed concrete quantum algorithms for measuring the characteristic function~\cite{xu2019probing}, a systematic investigation combined with digital signal processing theory is still lacking, which is the focus here. 

While such a concept of studying FCS by quantum computing is very natural, a systematic investigation, especially a concrete quantum algorithm for measuring the characteristic function, is still lacking, which is the focus here.

In this paper, we propose a quantum computing procedure for studying the full-counting statistics with digital quantum computers. A quantum algorithm is designed for calculating the characteristic functions with the Hadamard test. From the characteristic functions, the particle distribution as well as the cumulants can be derived by Fourier transformation. The process of sampling the characteristic function is similar to that of signal sampling, which involves sampling a continuous signal and then reconstructing the signal using the discrete sampled information. Thus, we analyze the complexity of this algorithm using digital signal processing (DSP) theory~\cite{rabiner1975theory,smith1997scientist,mitra2001digital,ifeachor2002digital,proakis2007digital,smith2013digital}. Moreover, we construct a quantum filter circuit for filtering out components in the quantum state with small weights of particle distribution, which can reduce the desired number of sampling points while retaining the accuracy of FCS. We analyze the time complexity of FCS as well as demonstrate the algorithm numerically with the mixed Ising model for evaluating the full-counting statistics of domain walls.

This paper is organized as follows. In Sec.~\ref{section 2}, we introduce the concept of FCS, then we will show how to implement it on a quantum computer and improve it with quantum filtering. Then, in Sec.~\ref{section 3}, we present numerical simulation results for spin chains to demonstrate the algorithm. Finally, we present the conclusion in Sec.~\ref{section 4}.
    
\section{full-counting statistics of particle distribution on digital quantum computer}\label{section 2}
	
In this section, we first introduce the full-counting statistics. Then we show how to implement it on quantum computer and improve it with quantum filtering.

\subsection{Full-counting statistics}
For a probability distribution $p(x)$, there will be a corresponding characteristic function $\widetilde{p}(k)$, by Fourier transform of $p(x)$. In the context of quantum mechanics, the characteristic function can be expressed as the expectation value of the operator $e^{-ik\hat{x}}$. Obviously, this function will contain all the information about the probability distribution, because we only need to perform an inverse Fourier transform on it to obtain the original distribution.
\begin{eqnarray}
\widetilde{p}(k)=\langle e^{-ik\hat{x}} \rangle=\int{p(x)e^{-ikx}dx}.
\end{eqnarray}
The characteristic function is also known as the moment generating function as we can obtain the moments of each order by expanding its exponential term in series.
\begin{eqnarray}
\widetilde{p}(k)=\sum_{n=0}^{\infty}\frac{(-ik)^n}{n!}\langle x^{n} \rangle,
\end{eqnarray}
where $\langle x^n \rangle$ is the $n$-order moment. We can use the moments to calculate the cumulants. For instance, the cumulants of the first three orders are the familiar mean, variance and skewness, respectively.
\begin{eqnarray}\label{Eq.(r0)}
\langle x \rangle _c&=&\langle x \rangle\ , \nonumber \\
\langle x^2 \rangle _c&=&\langle x^2 \rangle - \langle x \rangle^2\ , \nonumber \\
\langle x^3 \rangle_c &=&\langle x^3 \rangle - 3 \langle x^2 \rangle \langle x \rangle\ + 2 \langle x \rangle^2
\end{eqnarray}
where $\langle x^n \rangle _c$ is the $n$-order cumulant.
	
Here, we focus on the particle distribution where the variable takes the integer number. The operator for measuring particle number can usually be written in the form of a sum of local operators, so it can be effectively implemented on quantum computer.
\begin{eqnarray}\label{Eq.(r1)}
\hat{N}=\sum_{i}^{L}\hat{N_i}.
\end{eqnarray}

We use the Kronecker delta and the particle number operator $\hat{N}$ to construct a projection operator that can project quantum states onto the subspace with a given particle number $n$. By using the integral representation of the Kronecker delta, the projection operator can be conveniently written as
\begin{eqnarray}
\delta[\hat{N}-n]=\frac{1}{2\pi}\int_{-\pi}^{\pi}d\theta e^{i\theta (\hat{N}-n)}.
\end{eqnarray}

The particle distribution can be obtained by the expectation value of this operator.
\begin{eqnarray}
P(n)=\langle \delta[\hat{N}-n] \rangle,
\end{eqnarray}
where the angular bracket denotes the expectation value with respect to the state of the system. By introducing the Fourier transform representation, this probability distribution can also be written in the following form.
\begin{eqnarray}
P(n)=\frac{1}{2\pi}\int_{-\pi}^{\pi}d\theta \widetilde{P}(\theta) e^{-i\theta n},
\end{eqnarray}
where the characteristic function $\widetilde{P}(\theta)$ reads as
\begin{eqnarray}
\widetilde{P}(\theta)=\langle e^{i\theta \hat{N}} \rangle.
\end{eqnarray}

\subsection{Quantum algorithm for FCS}

\begin{figure}[b]
\includegraphics[width=\linewidth]{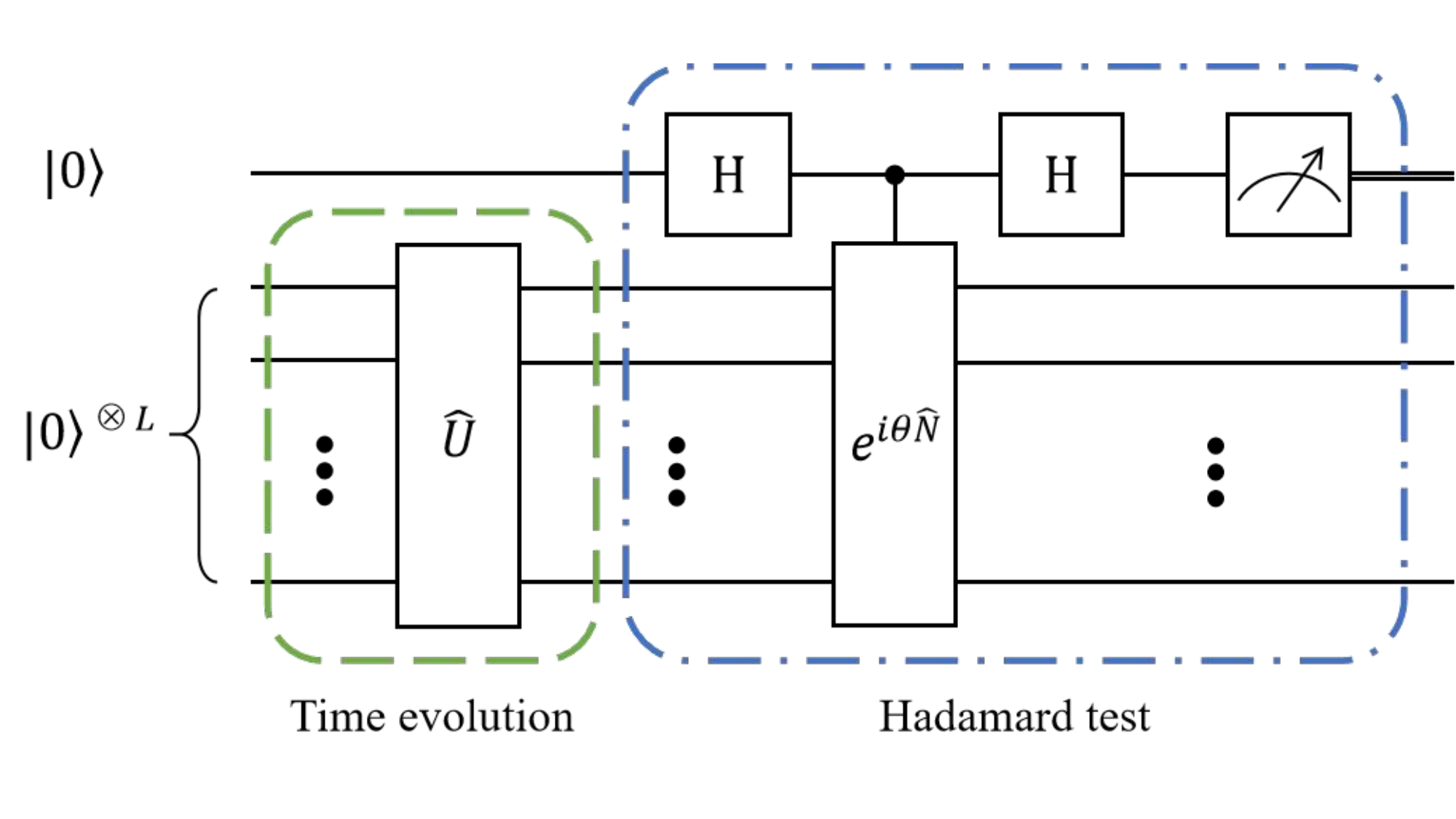}
\caption{Quantum circuit for evaluating the characteristic function. It prepares a state with $\hat{U}$. Then a Hadamard test is performed on the prepared state. The Hamiltonian of the controlled time-evolution is the particle number operator $\hat{N}$, and the evolution time is $-\theta$.\label{Figure 1}}
\end{figure}

The key step to this algorithm is to obtain the characteristic function $\widetilde{P}(\theta)$, which is the expectation value of the operator $e^{i\theta \hat{N}}$. Then we can obtain the particle distribution through discrete Fourier transform and cumulants through finite difference. Now we present the procedure to calculate the characteristic function $\widetilde{P}(\theta)$. First, the system is initialed in a product state of $|0\rangle$. Then, a unitary operator $\hat{U}$ is performed to prepare the quantum state,  
\begin{eqnarray}
|\psi(t)\rangle=\hat{U}|\psi_0\rangle=\hat{U}|0\rangle^{\otimes L}.
\end{eqnarray}
With an auxiliary qubit, the Hadamard test can obtain the expectation value of the operator $e^{i\theta \hat{N}}$. Herein, a controlled time-evolution unitary with the Hamiltonian $\hat{N}$ and the evolution time $-\theta$ is performed. Since the particle number operator $\hat{N}$ is a sum of local operators, the controlled time-evolution unitary can be effectively implemented. The quantum circuit is shown in Fig.~\ref{Figure 1}, which can get the real part of $\widetilde{P}(\theta)$. To get the imaginary part, we need to replace the Hadamard gate before measurement with $R_x(\frac{\pi}{2})$ gate.

\emph{Cumulants.}
The cumulants can be obtained as a linear combination of moments at different orders. It thus suffers to calculate the moments from the characteristic functions. The $n$-order moment can be written as $n$-order derivative of the moment generating function, 
\begin{equation}
\langle \hat{N}^n \rangle=\frac{\partial^n\widetilde{P}(\theta)}{\partial(i\theta)^n}\Bigg|_{\theta=0}.    
\end{equation}
For numeral evaluation, one can refer to finite difference scheme as an approximation for the derivative. It requires only a few sampling points to obtain the cumulants of low orders, such as the mean value, the variance and the skewness.
	
In essence, the finite difference is the finite truncation of the series expansion of the derivative. The accuracy of the finite difference formula depends on the number of truncated terms. The most commonly used finite difference formula can be written in the following form,
\begin{eqnarray}\label{Eq.(r2)}
&&\langle \hat{N} \rangle=\frac{\widetilde{P}(h)-\widetilde{P}(-h)}{2ih}+O(h^2), \nonumber \\ 
&&\langle \hat{N}^2 \rangle=\frac{\widetilde{P}(h)+\widetilde{P}(-h)-2\widetilde{P}(0)}{(ih)^2}+O(h^2), \nonumber \\
&&\langle \hat{N}^3 \rangle=\frac{\widetilde{P}(2h)-2\widetilde{P}(h)+2\widetilde{P}(-h)-\widetilde{P}(-2h)}{2(ih)^3}+O(h^2). \nonumber\\
\end{eqnarray}
A finite difference formula for arbitrary order derivative can be obtained by the recursive formula $\frac{\partial^{n+1}\widetilde{P}(\theta)}{\partial(i\theta)^{n+1}}=\frac{\partial}{\partial(i\theta)}\frac{\partial^n\widetilde{P}(\theta)}{\partial(i\theta)^n}$.

The error of the above finite difference is of the order of the square of the spacing $h$. The accuracy can be further improved by the Richardson extrapolation method~\cite{richardson1927viii}. The core idea is to use numerical approximations with different spacings to cancel out the lower-order errors, which can be briefly described as follows. Assuming that a certain function $g$ can be numerically approximated as $g(x_1)=f(x_1)+ax_1^p+bx_1^q+\cdots$, where $x_1$ is the spacing, $p$ is the order of the first error and $q$ is the order of the next error with $q>p$. Similarly, for another spacing $x_2$, we have $g(x_2)=f(x_2)+ax_2^p+bx_2^q+\cdots$.
To eliminate the $p$-order error in $f(x_2)$, we can make a subtract between $g(x_1)$ and $g(x_2)$,
\begin{eqnarray}\label{Eq.(r3)}
\begin{split}
&\frac{f(x_1)-rf(x_2)}{1-r}+\frac{b(x_1^q-rx_2^q)}{1-r}+\cdots\\
&=f'(x_1)+b'x_1^q+\cdots,
\end{split}
\end{eqnarray}
where $r=(\frac{x_1}{x_2})^p$, $b'=\frac{b[1-(\frac{x_2}{x_1})^{q-p}]}{1-r}$.  Now, we replace the original method $f(x_1)$ with a new approximation,
\begin{equation}
f'(x_1)=\frac{f(x_1)-rf(x_2)}{1-r}.  
\end{equation}
The new numerical result will have a higher order error term than the original form, thus improving the numerical accuracy. 
	
With the Richardson extrapolation method and setting $\frac{x_1}{x_2}=\frac{1}{2}$, 
a new finite difference formula can be proposed as follows, 
\begin{eqnarray}
\langle \hat{N} \rangle&=&-\frac{\widetilde{P}(2h)-\widetilde{P}(-2h)}{12ih}+\frac{2\widetilde{P}(h)-2\widetilde{P}(-h)}{3ih}
+O(h^4), \nonumber \\
\langle \hat{N}^2 \rangle&=&-\frac{\widetilde{P}(2h)+\widetilde{P}(-2h)}{12(ih)^2}+\frac{4\widetilde{P}(h)+4\widetilde{P}(-h)}{3(ih)^2} \nonumber \\
 &-&\frac{15\widetilde{P}(0)}{6(ih)^2}+O(h^4), \nonumber \\
\langle \hat{N}^3 \rangle&=&-\frac{\widetilde{P}(4h)-\widetilde{P}(-4h)}{48(ih)^3}+\frac{17\widetilde{P}(2h)-17\widetilde{P}(-2h)}{24(ih)^3}\nonumber\\
&-&\frac{-4\widetilde{P}(h)+4\widetilde{P}(-h)}{3(ih)^3}+O(h^4). 
\end{eqnarray}
The method for obtaining the finite difference formula of higher order derivatives is the same as above. We only need to substitute the higher order initial finite difference formula into Eq.~\eqref{Eq.(r3)}. Moreover, the Richardson extrapolation method can be iterated repeatedly until a satisfactory error term is obtained. Finally, we can use moments to obtain cumulants according to Eq.(\ref{Eq.(r0)}).
	
\emph{Particle distribution.}
The particle distribution can be obtained through the discrete Fourier transform.
\begin{eqnarray}
P(n)\approx\frac{1}{2\pi}\sum_{i}^{k}\Delta\theta\widetilde{P}(\theta_i) e^{-i\theta_i n},
\end{eqnarray}
where $\{\theta_i\}$ is a set of uniformly sampled points in $[-\pi,\pi]$, with a total number of $k$. $\Delta\theta=\frac{2\pi}{k}$ is the interval of the sampling. How large $k$ is required for lossless sampling? According to the signal sampling theorem~\cite{shannon1948mathematical}, we need to sample at twice the highest frequency of the sampled signal in order to reconstruct the signal without distortion. It is possible, however, to reduce the number of sampling points, as revealed below. 
    
\subsection{FCS with filtering}
\begin{figure}[b]
\includegraphics[width=\linewidth]{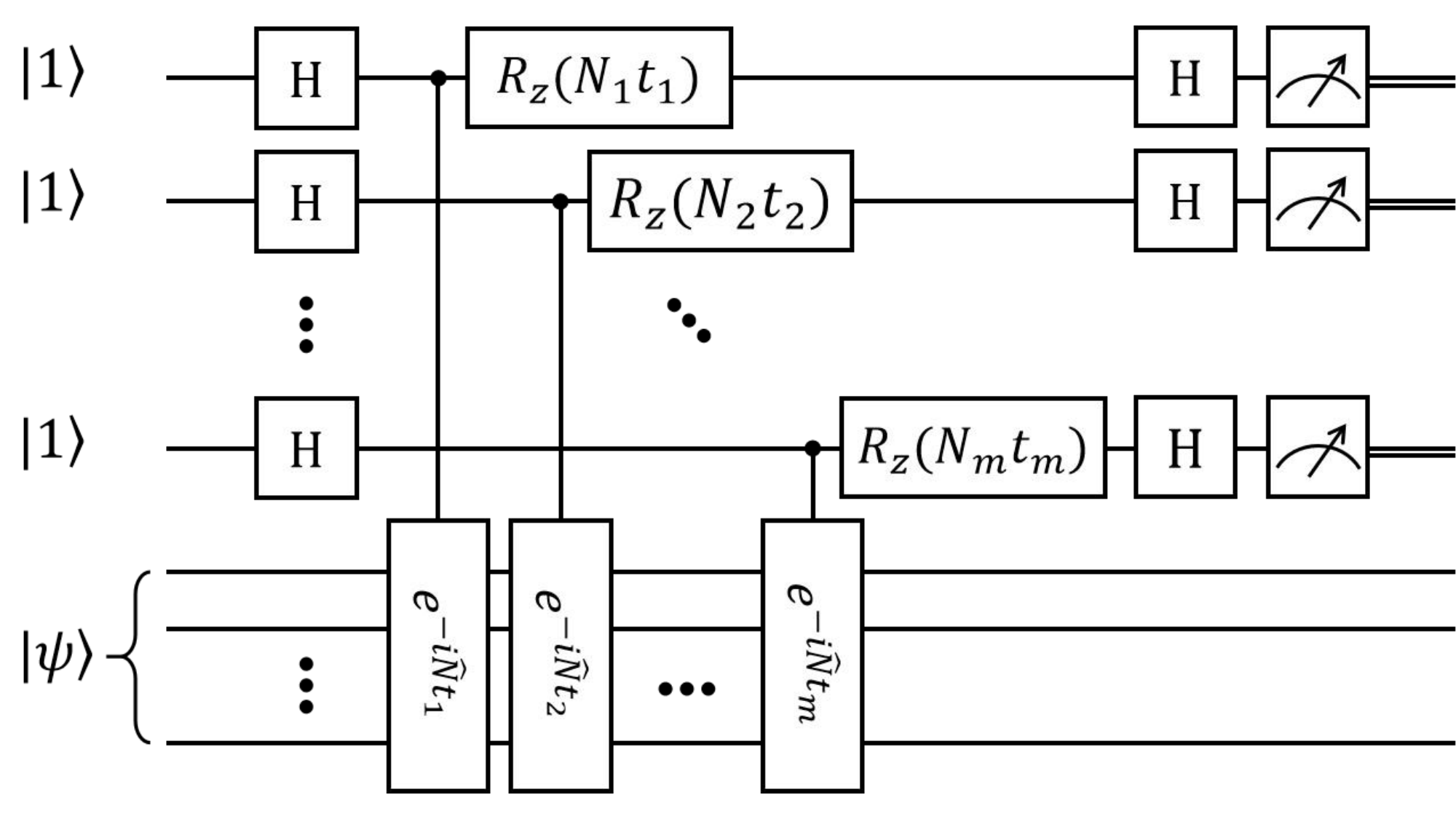}
\caption{Quantum circuit for quantum filtering.  Each auxiliary qubit is used as the control qubit for the controlled time evolution, with Hamiltonian being the particle number operator $\hat{N}$, and the evolution time being $t_m$. The phase rotation gate $R_z(N_mt_m)$ is performed on the $m$-th auxiliary qubit, where $N_m$ is the particle number that is desired to filter out.\label{Figure 2}}
\end{figure}

When we are only interested in the low-frequency part of the signal, this approach is obviously inefficient, because we spend extra resources to infer information about the high-frequency part. If we sample directly at a frequency lower than the sampling theorem, the low frequency signal reconstructed may be distorted significantly, which is known as signal aliasing~\cite{rabiner1975theory,smith1997scientist,mitra2001digital,ifeachor2002digital,proakis2007digital,smith2013digital}. For signal processing, an effective method is to add a low-pass filter before sampling to filter out the high-frequency part of the signal, and then use a lower sampling frequency than before to sample the filtered signal. Inspired by this, we propose a quantum filter circuit that can filter out components of certain particle numbers in a quantum state. The quantum filter circuit is a variant of the one proposed in Ref.~\cite{choi2021rodeo}. This process can be mainly divided into two steps.

\emph{Step 1: Quantum state filtering.}
The quantum circuit for illustration of quantum state filtering is shown in Fig.~\ref{Figure 2}. Firstly, the system qubits are initialized in a state $|\psi(t)\rangle$. Each auxiliary qubit is initialized in $|1\rangle$ and  is performed with a Hadamard operator. Afterward, a set of controlled Hamiltonian evolutions are applied, where the Hamiltonian is the particle number operator $\hat{N}$ and the evolution time series are $\{t_m\}$. The $m$-th auxiliary qubit serves as the control qubit for the Hamiltonian evolution with time $t_m$. Then, the phase rotation gate $R_z(N_mt_m)$ is performed on the $m$-th auxiliary qubit, where $N_m$ is the particle number desired to filter out. Then a Hadamard gate is applied to each auxiliary qubit. Finally, a projection to $|0\rangle$ for each auxiliary qubit marks the success of filtering. 
	
For demonstration, we will derive the case where only the component with particle number $N_1$ is filtered out. For a quantum state $|\psi\rangle=\sum_{i}c_i|\phi_i\rangle$, we can rewrite it by grouping $|\phi_i\rangle$ according to its particle number $N_k$.
\begin{eqnarray}
|\psi\rangle=\sum_{k}\sum_{i\in G_k}c_i|\phi_i\rangle,
\end{eqnarray}
where $G_k$ represents the set of subscripts for $\{|\phi_i\rangle\}$ with the particle number $N_k$. With $|\psi\rangle$ as the input state in Fig.~\ref{Figure 2}, the quantum state after projecting the auxiliary qubit to $|0\rangle$ will collapse into
\begin{eqnarray}
\sum_{k}\sum_{i\in G_k}\frac{c_i(\frac{1}{2}-\frac{1}{2}e^{-i(N_k-N_1)t_1})}{\sqrt{\sum_{j=i}^{2^L}|c_j(\frac{1}{2}-\frac{1}{2}e^{-i(n_j-N_1)t_1})|^2}}|0\rangle|\phi_i\rangle.
\end{eqnarray}
The coefficients of all $|\phi_i\rangle$ for $i\in G_1$ are equal to zero, verifying that the component in $|\psi\rangle$ where the particle number equals $N_1$ has been filtered out. The probability of successful filtering $P_f$ is the probability of the auxiliary qubit projected to $|0\rangle$.
\begin{eqnarray}\label{Eq.(r4)}
\begin{split}
P_f&=\sum_{k}\sum_{i\in G_k}|c_i(\frac{1}{2}-\frac{1}{2}e^{-i(N_k-N_1)t_1})|^2\\
&=\sum_{k}\sum_{i\in G_k}|c_i|^2\sin^2[(N_k-N_1)\frac{t_1}{2}].
\end{split}
\end{eqnarray}
From the above equation, we can see that the probability of successful filtering is determined by the parameter $t_1$, the filtered component $N_1$, and the probability distribution $\{|c_i|^2\}$ itself. To set a suitable parameter $t_1$ for adjusting the success probability, we need some prior knowledge about the distribution, such as the approximate position of the distribution center $N_c\in\{N_k\}$.  We can obtain a relatively high success probability by setting the parameter $t_1$ to the following form.
\begin{eqnarray}
t_1=
\begin{cases}
\frac{\pi}{|N_c-N_1|}, &\text{if $N_1\neq N_c$}\\ 
\frac{\pi}{|N_{c'}-N_1|}, &\text{if $N_1=N_c$} 
\end{cases}
\end{eqnarray}
where $N_{c'}\in\{N_k\}$ represents the point mostly close to $N_c$.

According to Eq.~\eqref{Eq.(r4)}, it can be seen that after setting the appropriate $t_1$, the success probability of filtering out the component with a smaller weight is higher than that of filtering out the component with a larger weight. It should be noted that the conclusion is not true for all $t_1$. For example, when $t_1$ is a small value, the success probability of any component will be very small.

\begin{table*}
\caption{\label{table 1}Comparison of resources used to obtain particle number distribution by FCS with and without quantum filtering. $L$ is the qubits number of the object system, $\epsilon$ is the precision of the results, $O(C_R)$ is the state preparation depth.}
\begin{ruledtabular}
\begin{tabular}{cccccc}
Algorithm & Auxiliary qubits number & Sampling points number & Circuit depth & Shots & Total runtime \\ \hline
without filtering & 1 & $O(L)$ & $O(C_R+L)$ & $O(\frac{1}{\epsilon^2})$ & $O((LC_R+L^2)\frac{1}{\epsilon^2})$ \\
with filtering & $m+1$ & $O(L-m)$ & $O(C_R+(m+1)L)$ & $O(\frac{1}{\epsilon^2})$ & $O((L-m)(C_R+(m+1)L)\frac{1}{\epsilon^2})$ \\
\end{tabular}
\end{ruledtabular}
\end{table*}

\emph{Step 2: Reconstruct the particle distribution.}
The particle distribution without filtering is $P(N=N_k)=\sum_{i\in G_k}|c_i|^2$. The distribution is changed with filtering which turns to be, 
\begin{eqnarray}\label{Eq.(r6)}
P'(N=N_k)&=&\sum_{i\in G_k}|c'_i|^2=\sum_{i\in G_k}g_k|c_i|^2\nonumber \\
&=& g_k\sum_{i\in G_k}|c_i|^2=g_k P(N=N_k),
\end{eqnarray}
where 
\[g_k=\frac{|\frac{1}{2}e^{-i(N_k-N_1)t_1}|^2}{\sum_{j=1}^{2^L}|c_j(\frac{1}{2}e^{-i(n_j-N_1)t_1})|^2}.\]
It is desirable to restore the original particle distribution $P(N=N_k)$ by evaluating the set of factors $\{g_k\}$. Note
the probability is changed for all $i\in G_k$ with the same factor $g_k$. Thus, we can select one $i\in G_k$ to measure the probabilities of projection to $|\phi_i\rangle$ with  filtering $P (i)$ and without filtering $P' (i)$. The factor $g_k$ then can be obtained as
\begin{eqnarray}\label{Eq.(r8)}
g_k=\frac{|c'_i|^2}{|c_i|^2}=\frac{P'(i)}{P (i)}.
\end{eqnarray}
Then we can get the desired particle distribution, 
\begin{eqnarray}\label{Eq.(r7)}
P(N=N_k)=\frac{P(i)}{P'(i)}P'(N=N_k).
\end{eqnarray}

The above procedure for filtering one component can be generalized for For multiple components. It needs more auxiliary qubits as shown in Fig.~\ref{Figure 2}, where each auxiliary qubit is used to filter out one component. The component is determined by the parameter $N_m$ of the phase rotation gate $R_z(N_mt_m)$. It can be verified that the quantum state after projecting all auxiliary qubits to $|0\rangle$ will collapse into
\begin{eqnarray}
\sum_{k}\sum_{i\in G_k}\frac{c_i\prod \limits_{m}(\frac{1}{2}-\frac{1}{2}e^{-i(N_k-N_m)t_m})}{\sqrt{\sum_{j=i}^{2^L}|c_j\prod \limits_{m}(\frac{1}{2}-\frac{1}{2}e^{-i(n_j-N_m)t_m})|^2}}|0\rangle^{\otimes m}|\phi_i\rangle.
\end{eqnarray}
Now, we complete the quantum state filtering. The probability of successful filtering $P_f$ is the probability that all auxiliary qubits are projected to $|0\rangle$.
\begin{eqnarray}
\begin{split}
P_f&=\sum_{k}\sum_{i\in G_k}|c_i\prod \limits_{m}(\frac{1}{2}-\frac{1}{2}e^{-i(N_k-N_m)t_m})|^2\\
&=\sum_{k}\sum_{i\in G_k}|c_i|^2\prod \limits_{m}\sin^2[(N_k-N_m)\frac{t_m}{2}].
\end{split}
\end{eqnarray}
Similar to the demonstration, the probability of successful filtering can be controlled by adjusting the parameter $t_m$. Therefore, the parameter $t_m$ needs to be set to the following form.
\begin{eqnarray}\label{Eq.(r5)}
t_m=
\begin{cases}
\frac{\pi}{|N_c-N_m|}, &\text{if $N_m\neq N_c$}\\ 
\frac{\pi}{|N_{c'}-N_m|}, &\text{if $N_m=N_c$} 
\end{cases}
\end{eqnarray}
where $N_{c'}\in\{N_k\}$ represents the point mostly close to $N_c$. The particle distribution after filtering $P'(N=N_k)$ still satisfies Eq.(\ref{Eq.(r6)}), where the factor $g_k$ should be modified as,
\begin{eqnarray}
g_k=\frac{\prod \limits_{m}|\frac{1}{2}e^{-i(N_k-N_m)t_m}|^2}{\sum_{j=1}^{2^L}|c_j\prod \limits_{m}(\frac{1}{2}e^{-i(n_j-N_m)t_m})|^2}.
\end{eqnarray}
Therefore, we can still perform some projective measurements and calculate the desired particle distribution according to Eq.(\ref{Eq.(r8)}) and Eq.(\ref{Eq.(r7)}).

\subsection{Analysis of complexity}
We give a comparison of complexity for FCS with and without quantum filtering, which is shown in Table~\ref{table 1}. The analysis is as follows.  Regarding the number of auxiliary qubits, FCS without filtering only requires one auxiliary qubit for the Hadamard test, while FCS with filtering requires additional auxiliary qubits for performing quantum filtering. In Table~\ref{table 1}, we assume that $m$ components of the quantum state are filtered out, so FCS with filtering requires $m+1$ auxiliary qubits. 

The number of sampling points is determined by the highest frequency of the signal according to the sampling theorem~\cite{shannon1948mathematical}, which corresponds to the maximum value of particle number of the quantum state. For FCS with filtering, we can see from Eq.~\eqref{Eq.(r1)} that the maximum particle number is $O(L)$. For FCS with filtering, the maximum value of the filtered state is $O(L-m)$ after $m$ components of large particle numbers are filtered out. Thus, the required sampling points number for both are $O(L)$ and $O(L-m)$, respectively.

The circuit depth has contributions from the state preparation, the Hadamard test, and the filtering. We set the circuit depth for the state preparation as $O(C_R)$, which depends on the problem under investigation. In the Hadamard test, it requires a controlled Hamiltonian evolution of the number operator $\hat{N}$ and the circuit depth is $O(L)$. For FCS with filtering, the circuit depth for the filtering should be accounted. There will be a series of controlled Hamiltonian evolution, for each the circuit depth is $O(L)$. Thus, it requires a circuit depth of $O(mL)$ for filtering out $m$ components. In total, the circuit depths for FCS without and with filtering is $O(C_R+L)$ and $O(C_R+(m+1)L)$, respectively. It should be emphasized that the filtered components only take a small weight and thus, the success probability for filtering is $O(1)$.  The number of shots for measurements with statistical errors $\epsilon$ is $O(1/\epsilon^2)$ for both. Finally, we give the total time complexity by taking all the above contributions into account. 
 
We may analyze the situation in that the filtering has an advantage by comparing the total time complexities. The advantage with filtering holds for $O(LC_R+L^2)>O((L-m)(C_R+(m+1)L))$. This can be reduced to two scenarios. One is $O(C_R)\le O(L^2)$ and $O(m)>O(L-C_R/L)$, which leads to $O(m)>O(L)$ and thus is not practical. The other is $O(C_R)>O(L^2)$ and $m>0$, which suggests that there will be an advantage with filtering once the circuit depth for preparing the quantum state is large enough.

\section{simulation results}\label{section 3}

In this section, we demonstrate the quantum algorithm using a model Hamiltonian. The numeral simulation is conducted using the open-source package \textit{ProjectQ}~\cite{steiger2018projectq,haner2018software}.

We consider the one dimensional mixed field Ising model (MFIM). This model can be described by the following Hamiltonian.
\begin{eqnarray}
\hat{H}=-J\sum_{i=1}^{L}[\sigma_i^z\sigma_i^{z+1}+h_x\sigma_i^x+h_z\sigma_i^z],
\end{eqnarray}
where $\sigma_i^\alpha$,$\alpha\in x,y,z$ are the Pauli matrices acting on the $i$th site, $i\in \{1,2,...,L\}$, $L$ is the length of the chain, $J$ is the Ising exchange of nearest neighbor spin $1/2$ and $h_{x/z}$ are the relative strengths of the transverse and longitudinal fields, respectively. The MFIM reduces to the transverse field Ising model(TFIM) at $h_z=0$, which can be exactly solved via the Jordan-Wigner transformation and describes free fermions. We consider periodic boundary conditions $\sigma_{L+1}=\sigma_1$ with an even $L$, for simplicity.  For TFIM, there are excitations of domain walls that correspond to configurations with opposite spins for two adjacent sites. The domain wall can also be considered as fermionic excitation after the Jordan-Wigner transformation. With an additional longitudinal field, there will be a confining potential between two domain walls. For later use, we define the operator measuring the domain wall number as,
\begin{eqnarray}
\hat{N}\equiv\frac{1}{2}\sum_{i=1}^{L}(1-\sigma_i^z\sigma_{i+1}^z).
\end{eqnarray}

To demonstrate our algorithm for studying the FCS of interacting systems, we are studying the number distribution of domain walls for MFIM after a Hamiltonian evolution with a time interval $t$ from an initial state $|0\rangle^{\otimes L}$. For all numeral simulations, we set $L=12$, $J=1$, $h_x=1$, $h_z=1$ and $t=1$, and compare the simulation results with those of exact diagonalization (ED).

\begin{figure}
\includegraphics[width=\linewidth]{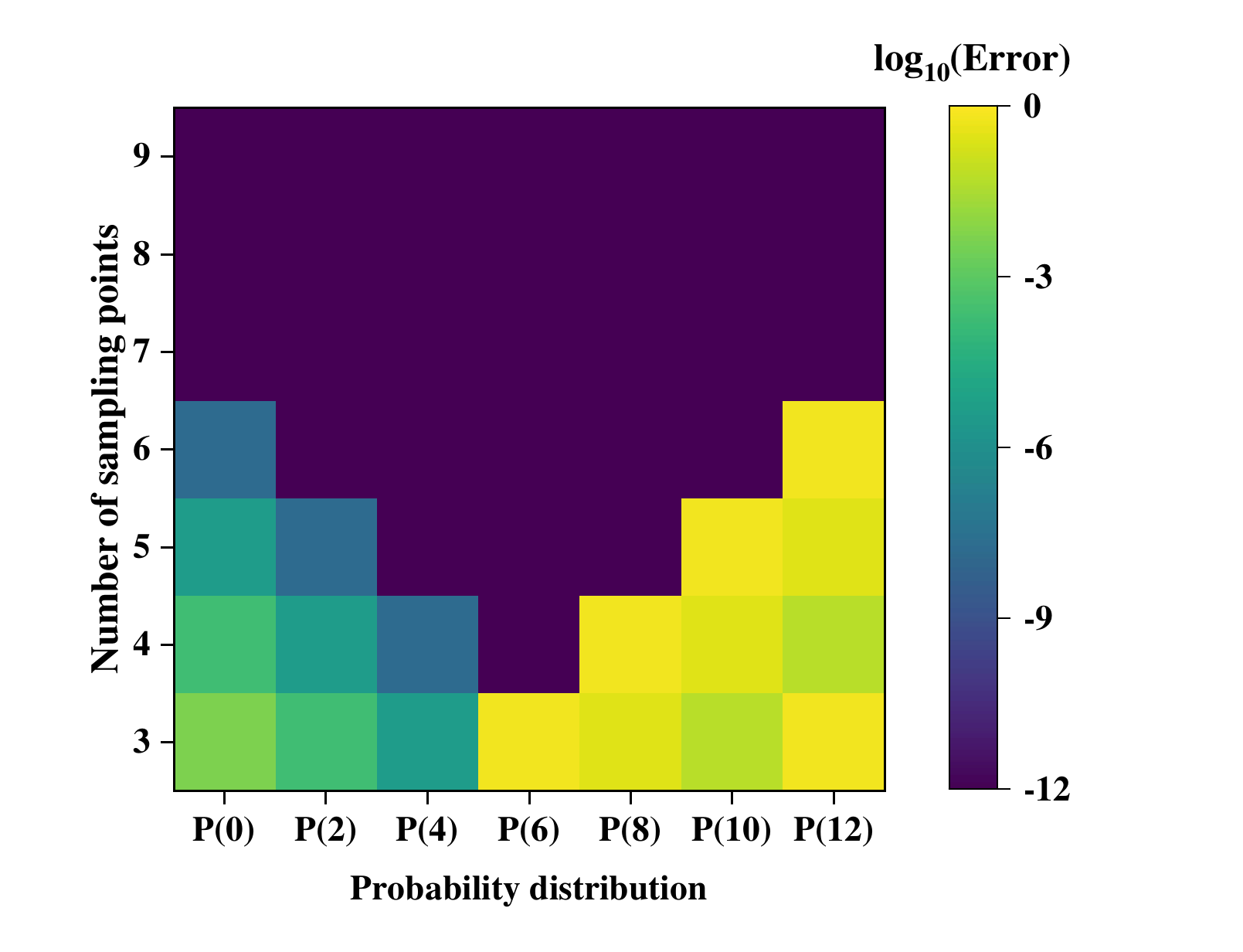}
\caption{The errors between results of FCS with different numbers of sampling points and results of exact diagonalization. Here we use the trace distance between two probabilistic distributions as a measure of error.}\label{Figure 3}
\end{figure}

\begin{figure}
\includegraphics[width=\linewidth]{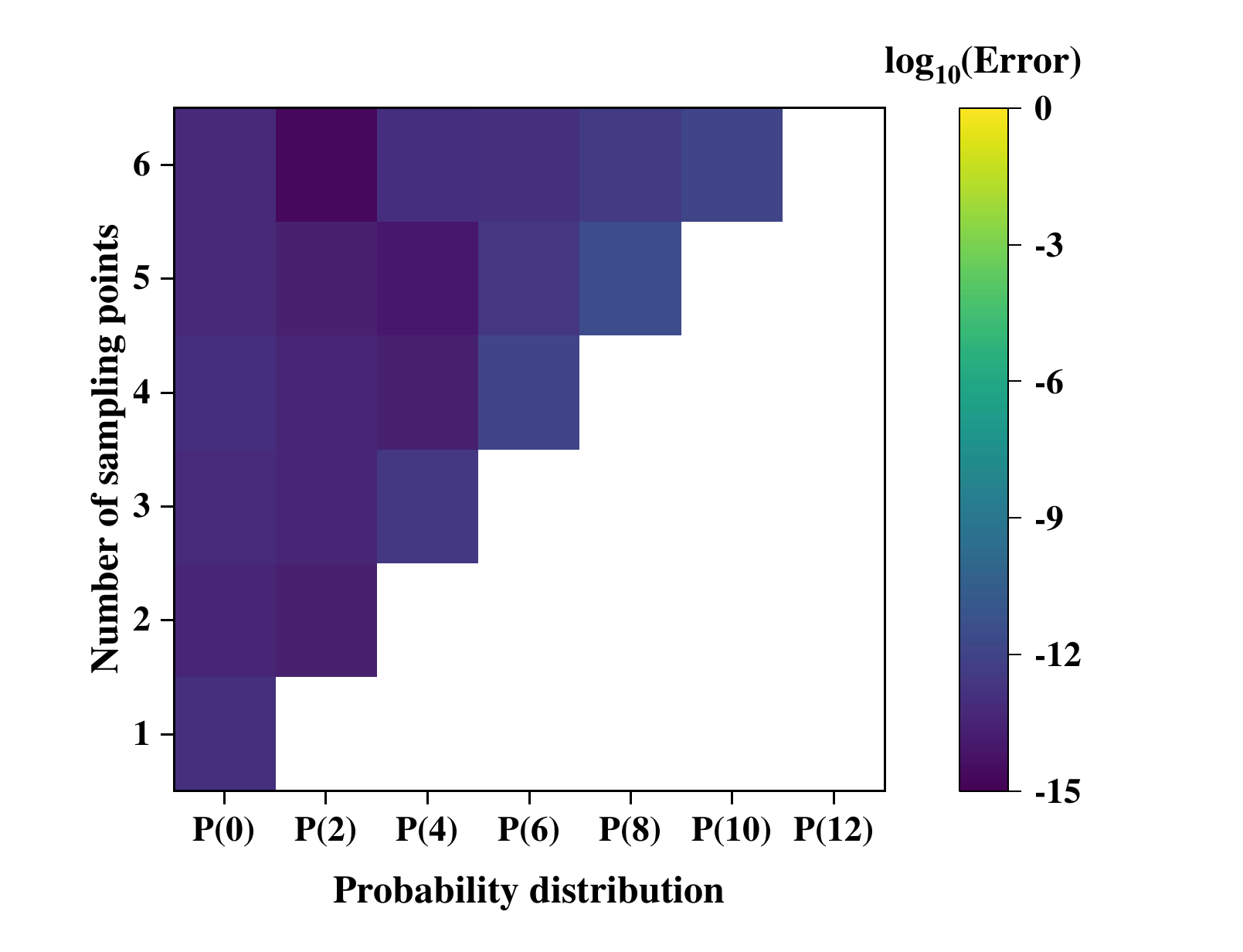}
\caption{The error between the results of FCS with quantum filtering and ED results. The components with large particle number have small weights and are filtered out, which are marked as a blank zone in the bottom right.\label{Figure 4}}
\end{figure}

\begin{figure}[b]
\includegraphics[width=\linewidth]{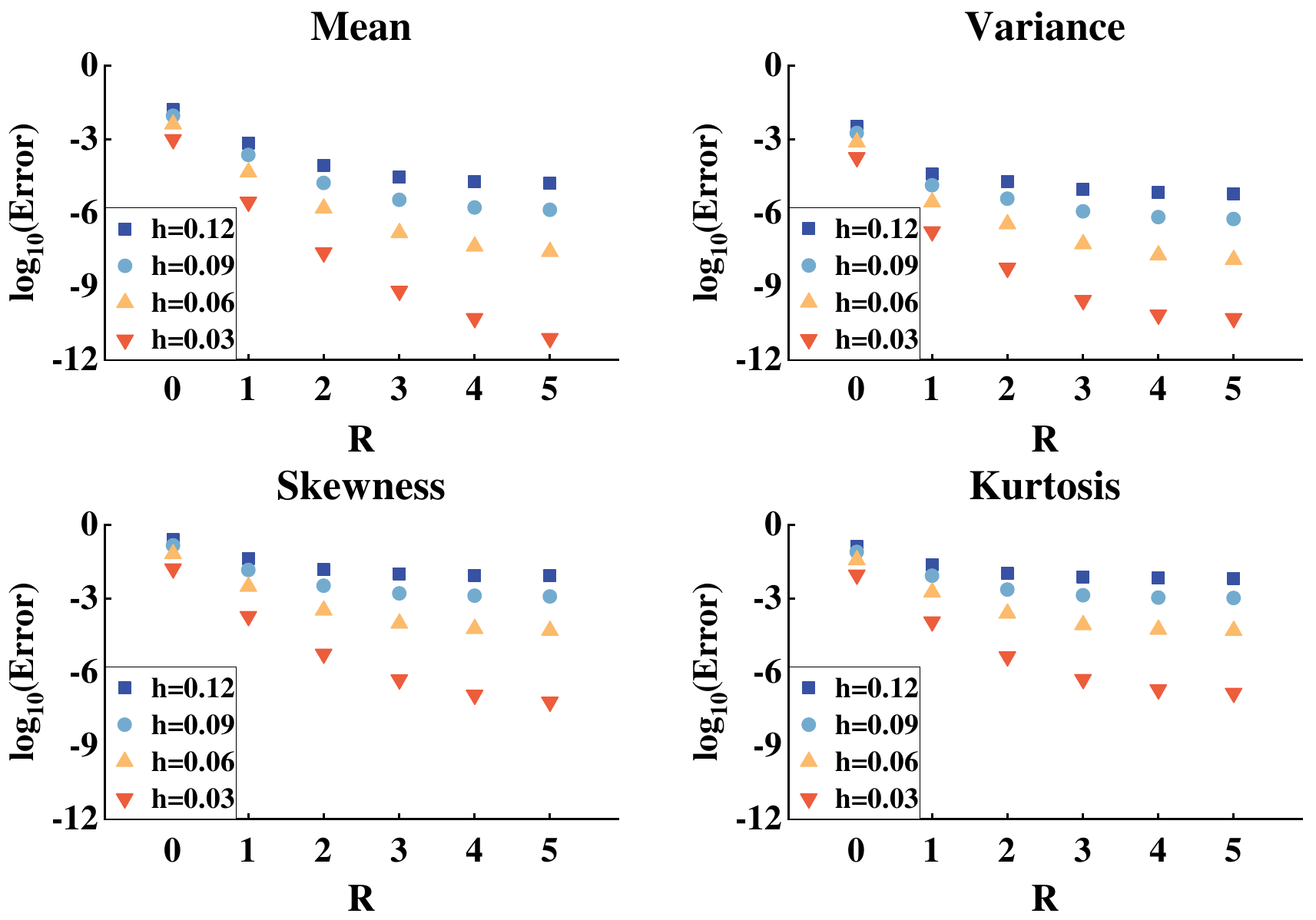}
\caption{The errors between results of the finite difference with Richardson extrapolation and exact diagonalization. $R$ represents the number of iterations of Richardson extrapolation and $h$ is the step size of the finite difference.}\label{Figure 5}
\end{figure}

Firstly, we present the results of FCS without quantum filtering. The errors between the simulation results of FCS with different numbers of sampling points and ED results is shown in Fig.~\ref{Figure 3}. We can clearly see that when the number of sampling points is lower than the minimum $\frac{L}{2}+1$, both the low-frequency and high-frequency parts of the spectrum will be aliasing. This will cause distortions for distributions of both small and large particle numbers, leading to a $V$-shape structure in the figure. For components in the intermediate frequency band, the results are of high accuracy. In other words, without quantum filtering, we can directly use the number of sampling points lower than $\frac{L}{2}+1$ to accurately obtain the particle distribution in the intermediate frequency regime. To retain the accuracy with a lower number of sampling points, it is necessary to refer to a quantum filter circuit, as shown below. We then present the results of FCS with quantum filtering. In our demonstration example, the number distribution is approximately a normal distribution, with the center located in the area with a small particle number. Therefore, we filter out the components with a large particle number to obtain the distribution of the remaining part more efficiently. The parameter $t_k$ is set according to Eq.(\ref{Eq.(r5)}), where $N_c=0$. The error between the simulation results of FCS with quantum filtering at different numbers of sampling points and ED results is shown in Fig.~\ref{Figure 4}. The blank zone in the bottom right indicates that these components have been filtered out by quantum filtering. From Fig.~\ref{Figure 4}, we can see that with filtering, the low-frequency distribution can be obtained with fewer sampling points. This is an improvement to the FCS algorithm. 
%\revise{It should be noted that we used a trick in the process of obtaining the numerical results of Fig.~\ref{Figure 3} and Fig.~\ref{Figure 4}. This trick is that we actually use the operator $\hat{N}/2$ to obtain the particle distribution. This can make the highest frequency smaller and the eigenvalues still be integers.}

Then, we show the results of cumulants obtained by the finite difference method. Errors between the results of finite difference   and the results of exact diagonalization are shown in Fig.~\ref{Figure 5}. Here, $h$ is the step size of the finite difference and $R$ represents the number of iterations of Richardson extrapolation, where $R=0$ corresponds to the initial finite difference formula in Eq.~\eqref{Eq.(r2)}. From Fig.~\ref{Figure 5}, we can see that as $R$ increases, the errors of the finite difference  gradually decrease. The simulation results clearly show the efficiency of Richardson extrapolation for improving the accuracy of using finite difference formula for evaluating the low-order cumulants.

\section{conclusion}\label{section 4}
In summary, we have proposed to perform the full-counting statistics by quantum computing which has an advantage over classical methods for generic interacting quantum systems. The algorithm relies on a quantum routine to measure the characteristic function and then extract the particle distribution and the cumulants by the discrete Fourier transform and the finite difference method, respectively. We have also used the Richardson extrapolation method to improve the accuracy of the finite difference results. Using the digital signal processing theory, we have analyzed the complexity of the algorithm, which involves the number of sampling points of the characteristic function. Moreover, we have constructed the quantum filter for filtering out components with a specific number of particles. We have demonstrated the effectiveness and the correctness of our algorithm. Our work provides an avenue for studying full-counting statistics of interacting systems with quantum computers.

\begin{acknowledgments}
This work was supported by the National Natural Science Foundation of China (Grant No. 12005065) and the Guangdong Basic and Applied Basic Research Fund (Grant No.2021A1515010317).
\end{acknowledgments}

\bibliography{FCS}
	
\end{document}